\documentclass[twoside,leqno,twocolumn]{article}

\usepackage[letterpaper]{geometry}

\usepackage{ltexpprt}
\usepackage[hidelinks]{hyperref}
\usepackage{adjustbox}

\usepackage{amssymb,times,enumerate,amsmath,graphicx,color,ifthen,amsfonts,url}
\usepackage{algorithm,algorithmic,stmaryrd,newfile,mathtools,bm}
\usepackage{paralist}
\usepackage{subfig}
\usepackage{graphicx}
\usepackage{setspace}

\usepackage{etoolbox}
\newcommand{\algorithmicdoinparallel}{\textbf{do in parallel}}
\makeatletter
\AtBeginEnvironment{algorithmic}{%
  \newcommand{\FORP}[2][default]{\ALC@it\algorithmicfor\ #2\ %
    \algorithmicdoinparallel\ALC@com{#1}\begin{ALC@for}}%
}
\makeatother

\makeatletter
\newcommand{\linebreakand}{%
  \end{@IEEEauthorhalign}
  \hfill\mbox{}\par
  \mbox{}\hfill\begin{@IEEEauthorhalign}
}
\makeatother

\usepackage[english]{babel}
\usepackage[utf8]{inputenc}
\usepackage{lipsum}

\usepackage{listings}
\definecolor{mygreen}{rgb}{0,0.2,0}
\definecolor{mygray}{rgb}{0.5,0.5,0.5}
\definecolor{mymauve}{rgb}{0.58,0,0.82}
\definecolor{mypurple}{rgb}{0.38,0,0.32}
\definecolor{myblue}{rgb}{0.1,0,0.32}
\newcommand{\costyle}{\footnotesize\ttfamily\bfseries}
\newcommand{\kwstyle}{\costyle\textcolor{myblue}}

\newcommand{\vcr}[1]{\bm{#1}}
\newcommand{\mat}[1]{\bm{#1}}

\newcommand{\rem}[1]{0}

\usepackage[normalem]{ulem}

\newcommand{\argmin}{\arg\!\min} 

\newcommand\blfootnote[1]{%
  \begingroup
  \renewcommand\thefootnote{}\footnote{#1}%
  \addtocounter{footnote}{-1}%
  \endgroup
}

\definecolor{dkgreen}{rgb}{0,0.6,0}
\definecolor{gray}{rgb}{0.5,0.5,0.5}
\definecolor{mauve}{rgb}{0.58,0,0.82}

\usepackage{etoolbox}

\begin{document}

\date{}
\title{Parallel Minimum Spanning Forest Computation using Sparse Matrix Kernels}
\author{Tim Baer \thanks{University of Illinois at Urbana-Champaign.} 
\and Raghavendra Kanakagiri \footnotemark[1] 
\and Edgar Solomonik  \footnotemark[1] 
}




\maketitle

\blfootnote{\\ We are grateful to Zhaoyu Wu and David (Yunxin) Zhang for early contributions to this project. This work used the Extreme Science and Engineering Discovery Environment (XSEDE), which is supported by National Science Foundation grant number ACI-1548562. Via XSEDE, the authors made use of the TACC Stampede2 supercomputer. The research was supported by the US NSF OAC via award No. 1942995.}

\begin{abstract}
Formulations of graph algorithms using sparse linear algebra have yielded highly scalable distributed algorithms for problems such as connectivity and shortest path computation. We develop the first formulation of the Awerbuch-Shiloach parallel minimum spanning forest (MSF) algorithm using linear algebra primitives. We introduce a multilinear kernel that operates on an adjacency matrix and two vectors. This kernel updates graph vertices by simultaneously using information from both adjacent edges and vertices. In addition, we explore optimizations to accelerate the shortcutting step in the Awerbuch-Shiloach algorithm. We implement this MSF algorithm with Cyclops, a distributed-memory library for generalized sparse tensor algebra. We analyze the parallel scalability of our implementation on the Stampede2 supercomputer. 
\end{abstract}


\section{Introduction}

Graph computations are ubiquitous in many disciplines with applicability to many real world problems. In recent years, there has been considerable interest in formulating graph algorithms via sparse linear algebra primitives. These primitives mask the underlying irregular communication patterns, lack of cache locality, and high synchronization costs of graph algorithms to achieve scalability. In this paper, we focus on the minimum spanning forest (MSF) problem. When the graph is connected, the minimum spanning forest is a minimum spanning tree (MST). MST has many practical applications including network design for computers, transportation, telecommunication, and electrical grids \cite{mst_survey, history_mst}. Approximation algorithms for several problems including traveling salesman, maximum flow, and weighted perfect matching invoke the computation of MST as subroutines.

Given an undirected weighted graph, a MSF is a subset of edges that connects all the vertices in each connected component with the minimum possible total edge weight. 
The minimum spanning forest is unique if each edge has a distinct weight. 
Boruvka, Prim, and Kruskal each proposed what has become three classic MSF algorithms in the literature. 
MSF algorithms generally rely on the cut property of minimum spanning trees: for any subset of vertices $S$, the minimum-weight edge with one endpoint in $S$ and one endpoint not in $S$ belongs to all minimum spanning trees. We call such an edge a minimum weight outgoing edge.

Parallel MSF algorithms also have a long history. 
Among the classical algorithms, Boruvka's exhibits a high degree of parallelism. 
Many parallel MSF algorithms including Boruvka's defines disjoint subgraphs of the MSF in some way and joins them together with minimum weight outgoing edges. Awerbuch and Shiloach (AS) propose a MSF algorithm that has $O(\log n)$ depth on $n+m$ processors, where $n$ is the number of vertices and $m$ is the number of edges in the graph \cite{as}. In this work, we develop a distributed-memory implementation of the AS algorithm based on sparse matrix algebra.

Many recent proposals on graph algorithms targeting scalability and performance in distributed-memory implementations use linear algebraic primitives \cite{lacc, fastsv, bc}.
The plurality of algebraic graph algorithm libraries~\cite{suitesparse, kokkos, graphulo} has motivated standardization efforts such as the GraphBLAS~\cite{graphblas}.
The Combinatorial BLAS (CombBLAS) library~\cite{combblas} has been one of the first to leverage sparse matrix--vector and matrix--matrix primitives to implement graph algorithms.
Their implementation of the Awerbuch-Shiloach connectivity algorithm as part of LACC~\cite{lacc} is perhaps the most closely related work to ours.
We leverage the Cyclops library for our algebraic MSF implementation, which has previously been used to implement betweenness centrality using sparse matrix multiplication~\cite{bc}.
We overview the basics of algebraic graph algorithms, and describe the additional challenges in developing an algebraic implementation for the Awerbuch-Shiloach MSF algorithm relative to connectivity in Section~\ref{sec:background}.

We propose a formulation of the AS MSF algorithm in terms of sparse matrix primitives in Section \ref{sec:algebraic_msf}. 
To achieve this, we introduce a multilinear kernel that updates graph vertices by simultaneously using information from both adjacent edges and vertices.
This kernel is different from the matrix-vector-product operations proposed in the existing literature including those in GraphBLAS~\cite{graphblas}.
The multilinear kernel permits an algebraic implementation of the AS MSF algorithm that does not incur overhead in updating the adjacency matrix (for the AS connectivity algorithm, matrix--vector products and vector operations suffice).
Further, we propose new optimizations for the shortcutting step in the AS algorithm, which transforms MSF trees into stars.
The optimized algorithm shortcuts all trees into stars (height-1 trees) using one round of communication.
We provide a cost analysis of the algorithm and optimizations in Section~\ref{sec:parallel_analysis}.
We then contrast our work to other related efforts in Section \ref{sec:related_work}.

We implement this as part of a general distributed tensor library, Cyclops Tensor Framework (CTF) \cite{ctf} that supports a variety of generalized vector/matrix/tensor operations and some all-at-once multi-tensor contraction kernels~\cite{tttp}. 
In Section \ref{sec:evaluation}, the algebraic formulation of MSF using the multilinear kernel coupled with our optimizations achieves excellent strong and weak scaling for both synthetic and real world graphs. These real world graphs include some of the largest available graphs in the SNAP \cite{snapnets} and SuiteSparse datasets \cite{tamu}.


Overall, our paper makes the following novel contributions:
\begin{itemize}
    \setlength\itemsep{0.05em}
    \item we propose a new algebraic primitive that enables an efficient parallel implementation of the AS MSF algorithm~\cite{as},
    \item we provide the first distributed-memory implementation of the AS algorithm (the LACC implementation of the Awerbuch-Shiloach algorithm for connectivity~\cite{as} is the closest related work~\cite{lacc}),
    \item we propose a new optimization to the shortcutting procedure, which performs communication for all pointer-chasing rounds in a single stage,
    \item we demonstrate scalability of our algebraic implementation of the AS algorithm on up to 256 nodes (16K cores) of Stampede2 on graphs with up to 11 billion edges and 183 million vertices.
\end{itemize}




\section{Background} \label{sec:background}
We consider the case of an undirected weighted graph $G = (V, E, w:E\to \mathbb{R})$ with $n$ vertices, $m$ edges, and distinct edge weights. We label the vertices $\{1, \ldots, n\}$. A tree is an undirected graph in which any two vertices are connected by exactly one path. A directed rooted tree is a tree in which a single vertex is designated as the root and the edges of the tree are oriented toward the root. A star is a directed rooted tree of height at most $1$. A forest is a disjoint union of trees. A directed rooted forest is a disjoint union of directed rooted trees. An outgoing edge from a tree $T$ is an edge $(i,j)$ such that $i\in T$ and $j\notin T$. An outgoing edge from a vertex $i$ belonging to a tree $T$ is an outgoing edge from $T$ that is adjacent to $i$. A minimum outgoing edge from a tree is an outgoing edge with the smallest weight. A minimum outgoing edge from a vertex is an outgoing edge from that vertex with the smallest weight. The adjacency matrix $\mat{A}\in \mathbb{R}^{n\times n}$ of graph $G$ is defined by $a_{ij} = w(i,j)$ if $(i,j)\in E$ and $\infty$ otherwise.

\subsection{Basic Algebraic Structures}
~\\

\textbf{Monoids: } A monoid $(\mathbb{S}, \oplus)$ is a set $\mathbb{S}$ equipped with an associative binary operation $\oplus:\mathbb{S}\times \mathbb{S} \to \mathbb{S}$ called addition and an identity element. 

\textbf{Semirings: } A semiring is an extension of a monoid to two binary operations. Formally, a semiring $(\mathbb{S}, \oplus, \otimes)$ is a set $\mathbb{S}$ equipped with binary operations $\oplus,\otimes:\mathbb{S}\times \mathbb{S} \to \mathbb{S}$ called addition and multiplication respectively, satisfying: 
(i) additive associativity, (ii) additive commutativity, (iii) multiplicative associativity, and (iv) left and right distributivity.

These conditions imply the existence of additive and multiplicative identities for semirings. A simple example of a monoid is the set of binary strings equipped with a concatenation operation $(\{0,1\}^*, \cdot)$. The usual semiring in arithmetic is $(\mathbb{R}, +, *)$. Note that semirings do not guarantee the existence of additive inverses, so fast matrix multiplication algorithms like Strassen's algorithm~\cite{strassen} may not apply.



\subsection{Algebraic Graph Algorithms}
The formalism of algebraic structures such as monoids and semirings permit simple yet expressive graph algorithms. The key connection to linear algebra is the adjacency matrix representation of a graph. Many graph algorithms can be rewritten in terms of matrix-vector and matrix-matrix multiplications with the adjacency matrix on a certain monoid or semiring. 
Specialized sparse matrix-vector (SpMV) and sparse matrix-matrix multiplication (SpMSpM) algorithms have been designed with asymptotic complexity depending on $nnz$, the number of nonzero entries in the matrix. In particular, a SpMV may be computed with $O(nnz)$ floating point additions and multiplications.

As a simple example of an algebraic graph algorithm, we review a linear algebraic interpretation of the Bellman-Ford algorithm for single source shortest paths (SSSP) based on discussion from \cite{graph_algs_lang}. Formally, given a graph and a starting vertex $s$, the SSSP problem is to compute the length of a shortest path from $s$ to $j$ for all $j\in V$. The Bellman-Ford algorithm can be derived with dynamic programming on the number of hops on a shortest path from $s$ to $j$ for all $j\in V$. We store tentative shortest path distances in $\bm{d}^{(\ell)}\in \mathbb{R}^{n}$, where $d^{(\ell)}_{j}$ is the shortest path distance from $s$ to $j$ among paths with at most $\ell$ hops. The Bellman-Ford algorithm is based on the idea of \textit{edge relaxations}: we call an edge $(k,j)$ tense if $d^{(\ell)}_j > d^{(\ell)}_k + w(k,j)$. At each iteration, we \textit{relax} all tense edges, meaning that we update $d^{(\ell)}_j \gets d^{(\ell)}_k + w(k,j)$ if $(k,j)$ is tense. After $n-1$ iterations, $\bm{d}^{(n-1)}$ converges to SSSP.

%
%

\textbf{Algorithm:} Now, consider an interpretation of the Bellman-Ford algorithm that implements each edge relaxation with a SpMV of $\mat{A}$ and a row-vector $\bm{d}^{(\ell)}$. Instead of the usual semiring, we perform operations on $\mathbb{R}$ over the tropical semiring $(\mathbb{R},\oplus,\otimes)$ where $\oplus = \min$ and $\otimes = +$. We initialize $d^{(0)}_j$ to $0$ if $j=s$ and $\infty$ otherwise. We iteratively compute $\bm{d}^{(\ell+1)} \gets \bm{d}^{(\ell)} \mat{A}$ over the tropical semiring. After $n-1$ iterations, $d^{(n-1)}_j$ stores the length of a shortest path from $s$ to $j$ for all $j\in V$. Since we perform $n-1$ SpMVs with the adjacency matrix, the run time is $O(nm)$.

\textbf{Intuition:} Consider updating the tentative distances for a fixed $j$, $d^{(\ell+1)}_j \gets (\bm{d}^{(\ell)} \mat{A})_j = \bigoplus_{k} d^{(\ell)}_k \otimes  a_{kj}$. Replacing the generic semiring addition and multiplication symbols with $\min$ and $+$ respectively yields $d^{(\ell + 1)}_j \gets \min_{k} \{d^{(\ell)}_{k} + a_{kj}\}$. By introspection, we can interpret this expression as an edge relaxation.

\subsection{Awerbuch-Shiloach Algorithm} \label{ssec:as}
~\\
Awerbuch and Shiloach~\cite{as} (AS) provide a classic parallel algorithm for computing the minimum spanning forest of an undirected graph with distinct edge weights. In addition to the graph itself, the algorithm maintains a parent forest of directed rooted trees, each intuitively representing a part of the minimum spanning forest that has already been discovered. The algorithm grows the minimum spanning forest by joining trees together with minimum outgoing edges. The algorithm only computes minimum outgoing edges for trees that are stars. In such cases, we may find these edges with work proportional to the number of member vertices and in $O(1)$ depth.

We represent the minimum spanning forest with a set of edges $F$. In addition, we represent the parent forest with a parent vector $\bm{p}\in V^n$, where $p_i$ stores the parent of vertex $i$. In the first iteration, $F \gets \varnothing$, and the parent forest consists of $n$ isolated vertices, each with a self-loop. At each iteration, the algorithm computes the minimum outgoing edge $(i,j)$ for each star in parallel. We then join vertex $i$'s parent to vertex $j$'s parent and add edge $(i,j)$ to $F$, while taking care to prevent cycles in the parent forest. Next, we shortcut trees to reduce their height and possibly create many stars for the next iteration. The algorithm terminates when no more trees in the parent forest can be joined. If the input graph is connected, the algorithm will terminate when the parent forest converges to a single connected component and $F$ is the minimum spanning tree.

\begin{figure*}[t]
\centering
\subfloat[Star hooking: star roots $2$ and $7$ hook with their minimum outgoing edges. The tree rooted at $4$ is not a star so it does not hook.]{\includegraphics[width = 1.6in]{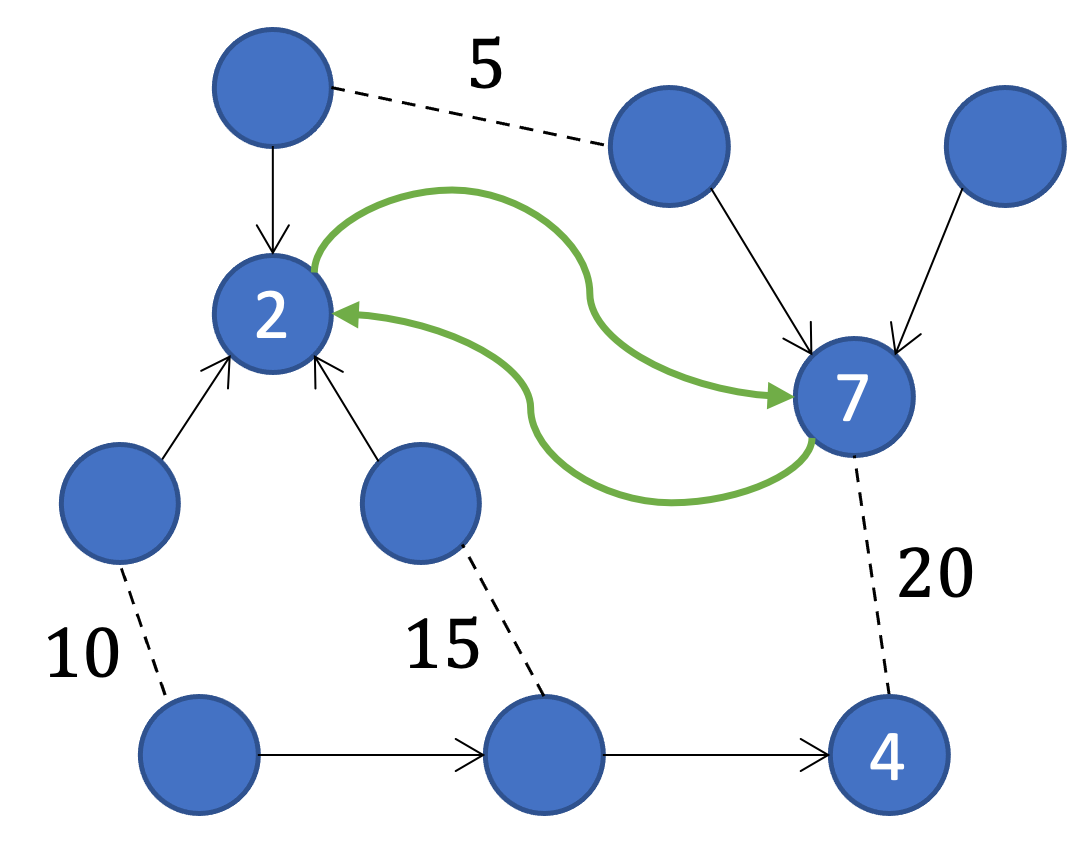}} 
\hspace{40pt}
\subfloat[Tie breaking: breaks the cycle introduced in star hooking by setting $2$ to be its own parent.]{\includegraphics[width = 1.6in]{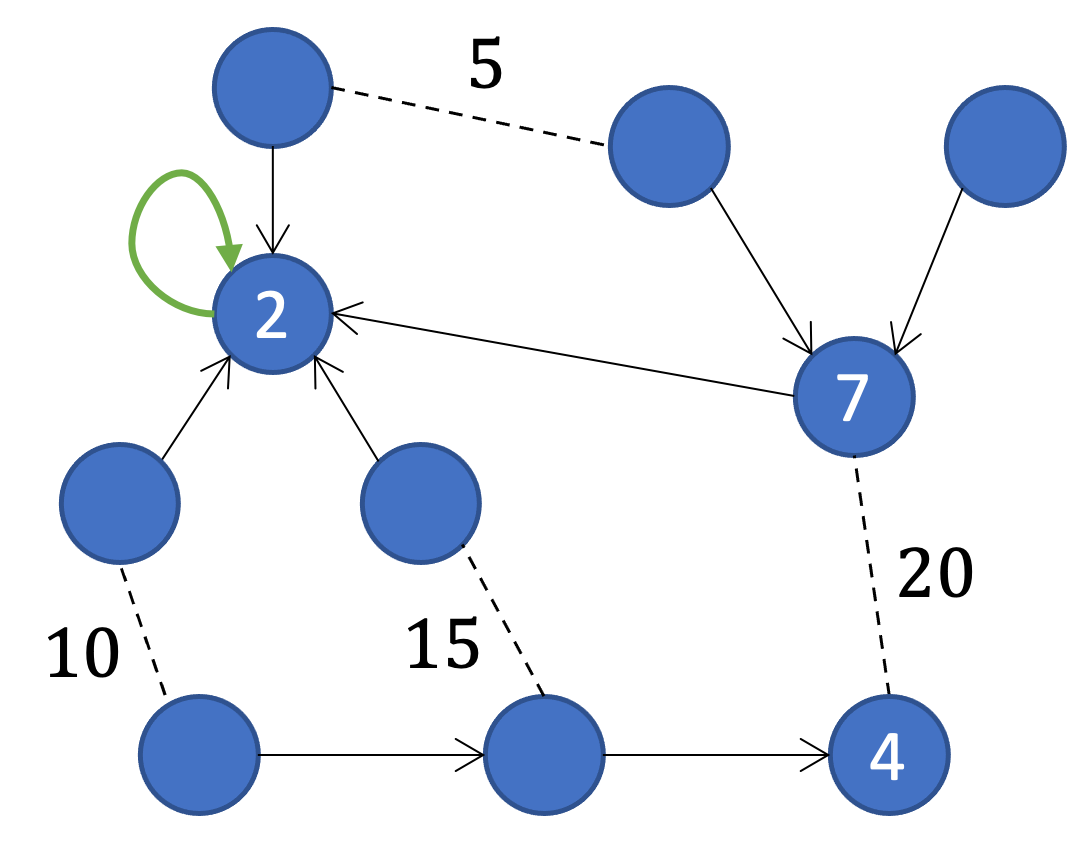}}
\hspace{40pt}
\subfloat[Shortcutting: changes each vertex's parent to the parent of its parent.]{\includegraphics[width = 1.6in]{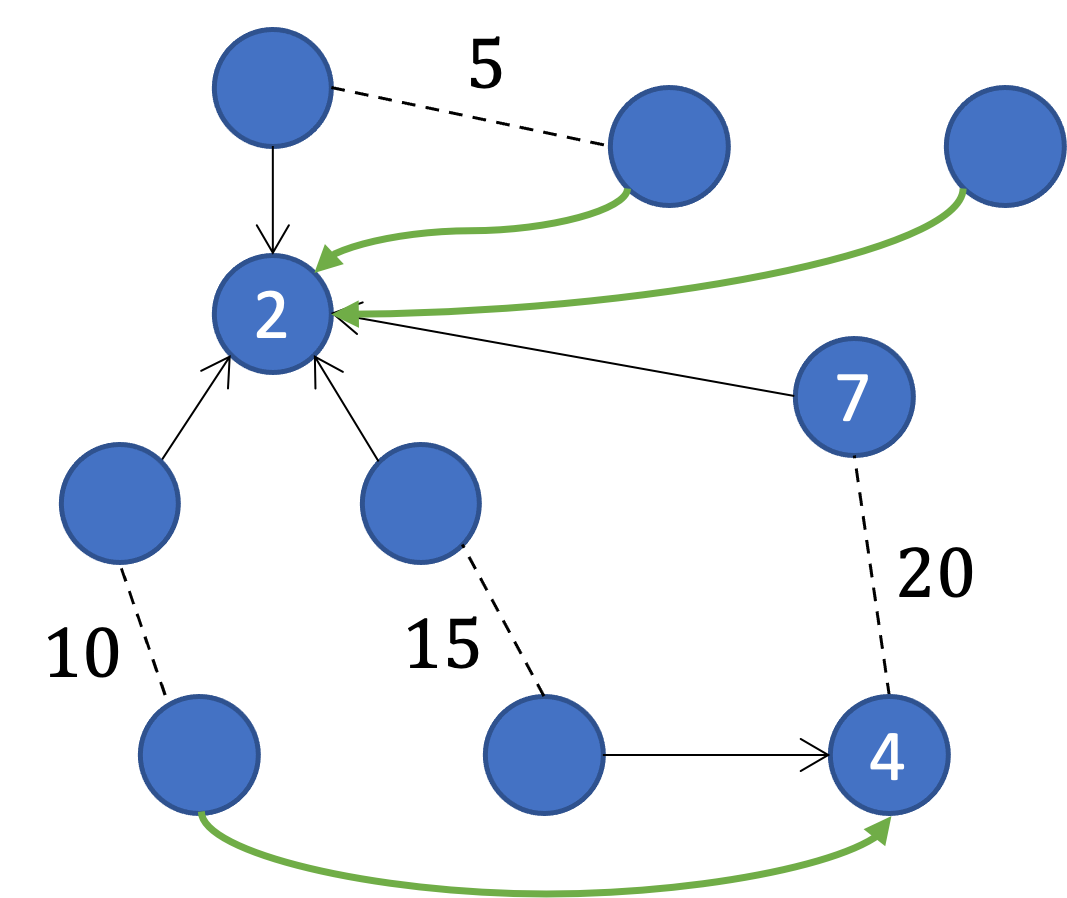}}
\caption{AS algorithm steps where dotted black lines are outgoing edges, solid black lines are parent pointers, and solid green lines are the updated parent pointers.}
\label{fig:as}
\vspace{-0.1in}
\end{figure*}

Let us assume that we have access to a routine that decides whether a given vertex belongs to a star. In summary, the algorithm performs three steps until convergence of the parent vector $\bm{p}$. \textit{Star hooking} joins two trees to grow the MSF. \textit{Tie breaking} breaks cycles by detecting and removing star hookings that create cycles. \textit{Shortcutting} reduces the height of trees by a factor of nearly two. We outline the AS algorithm in Algorithm \ref{alg:as}. Whenever an edge is used in star hooking, we add it to the MSF.
\begin{algorithm}[H]
\caption{Awerbuch-Shiloach}
\label{alg:as}
\begin{algorithmic}[1]
    \REQUIRE{$G=(V,E,w:E\to \mathbb{R})$ where $V=\{1,\dots,n\}$, is an undirected graph with distinct edge weights.}
    \STATE{Let $\vcr{p}^\text{old} = \begin{bmatrix} 0 & \cdots & 0 \end{bmatrix}^T$ be the old parent vector.}
    \STATE{Let $\vcr{p} = \begin{bmatrix} 1 & \cdots & n \end{bmatrix}^T$ be the parent vector.}
    \WHILE{$\bm{p}\neq \bm{p}^{old}$}
    \FOR[star hooking]{each star root $i$}
        \STATE{$(i,j) \gets \text{minimum outgoing edge from } i's \text{ star}$}
        \STATE{$p_i\gets p_j$}
    \ENDFOR
    \FOR[tie breaking]{each star root $i$}
        \IF{$i<p_i \text{ and } i=p_{p_i}$}
            \STATE{$p_i\gets i$}
        \ENDIF
    \ENDFOR
    \FOR[shortcutting]{each vertex $i$}
        \IF{$i$ does not belong to a star}
            \STATE{$p_i\gets p_{p_i}$}
        \ENDIF
    \ENDFOR
    \ENDWHILE
\end{algorithmic}
\end{algorithm}
We visualize these steps in Figure \ref{fig:as}. AS show that the sum of the heights of all the trees in the forest decreases by a factor of at least $3/2$ each iteration, resulting in convergence after $\log_{3/2} n$ iterations.

In \cite{as}, AS first present a parallel algorithm for computing the connected components (CC) of a graph that uses modified hooking and tie breaking steps, while reusing the shortcutting and starcheck steps from the MSF algorithm. For CC, it is sufficient to hook with any outgoing edge from a star instead of only the minimum such edge for the MSF. Computing the minimum outgoing edge from a star may not be expressed as a relaxation of the edges of a subset of vertices.

\textbf{Star hooking:} We consider the PRAM parallelization of a routine that computes the minimum outgoing edge from each star. In particular, we assume the concurrent read concurrent write (CRCW) PRAM with the conflict resolution that the write with the minimum value succeeds. Consider the edge $(i,j)\in E$ that is assigned to some processor. Suppose that the vertex $i$ belongs to a star. Then, the edge $(i,j)$ is outgoing from $i$'s star if $p_i\neq p_j$. Notice that the condition that vertex $i$ belongs to a star is necessary: if $i$ does not belong to a star, $(i,j)$ might not an outgoing edge but still have $p_i \neq p_j$. The processor that owns edge $(i,j)$ continues only if vertex $i$ belongs to a star. Next, the processor reads $p_i$ and $p_j$ to decide whether $p_i\neq p_j$. If $p_i\neq p_j$, then the processor writes $r_{p_i}\gets (i,j)$. We have computed the minimum outgoing edge $r_{p_i}$ from star root $p_i$ with $r_{p_i}\gets \argmin_{(i,j)\in E_i} w(i,j)$ where $E_i\subseteq E$ is the set of outgoing edges from vertex $i$. We see that this routine has work $O(m)$ and depth $O(1)$. By Brent's scheduling principle, we can compute this routine with $p$ processors in $O(m/p)$ steps. For weaker PRAM models without concurrent write or minimum conflict resolution \cite{pram_models}, we may simulate the update to $r_{p_i}$ with a slowdown of $O(\log p)$.

\textbf{Tie breaking:} Step $(ii)$ detects and breaks cycles of length $2$. We show that after star hooking, there can not exist cycles of length greater than $2$. Suppose for contradiction that there exists a cycle of length $k>2$. Given that the edge weights of the input graph are distinct, some edge in the cycle has the largest weight. Since the input graph is undirected and the edges form a cycle of length $k$, then this edge is not the minimum outgoing edge from any star. We see that step $(ii)$ is sufficient to remove any cycles introduced by star hooking.

\textbf{Starcheck:} Before each of the three steps, the algorithm requires vertices to determine whether or not they belong to a star.
First, each vertex checks $p_i=p_{p_i}$ to decide whether its parent is a root.
If not, the vertex does not belong to a star ($s_i=\text{False}$) and the vertex informs its parent's parent that the tree is not a star $(s_{p_{p_i}} = \text{False})$.
After this step, it suffices for each vertex that has not yet determined that it is not in a star 
to query their parent to determine whether they belong to a star (if $s_i=\text{True}, s_i=s_{p_i}$).
\subsection{Algebraic Connectivity Algorithms}
Given a MSF, each tree corresponds to a connected component of the original graph. 
In \cite{as}, AS also present a parallel algorithm for connectivity based on the Shiloach-Vishkin algorithm~\cite{sv}. Instead of hooking only with the minimum outgoing edge from a star, their connectivity variant hooks with \textit{any} outgoing edge. In addition, AS replace the tie breaking step with two different kinds of hooking: conditional and unconditional. The shortcutting and starcheck steps remain the same across their MSF and CC algorithms. In this section, we review recent work \cite{lacc,fastsv} on expressing the connectivity algorithm with linear algebra operations.

For conditional hooking, star roots are only allowed to hook onto other trees with a smaller parent index. Note that while this condition avoids cycles, it leads to $O(n)$ depth without the unconditional hooking step. In \cite{lacc}, Azad and Buluc rewrite the conditional hooking step as follows,
\begin{align*}
  & p^h_i \gets
  \begin{cases}
    \min_j \{p_j \} & : i \text{ belongs to a star},\\
    0 & : \text{ otherwise},
  \end{cases}\\
  & p_{p_i} \underset{\min}{\gets}
  \begin{cases}
    p^h_i & : p^h_i < p_i,\\
    p_i & : \text{otherwise}.
  \end{cases}
\end{align*}
First, we scan the neighbors of each vertex that belongs to a star and store the parent of a neighbor with the smallest index in $\bm{p}^h$. This step is implemented with a SpMV-like expression. Next, we only allow hooking onto other trees with a smaller parent index. The unconditional hooking step follows similarly, except without the check that $p^h_i < p_i$. The shortcutting and star checking steps can be implemented with reads/writes to vectors.

Performing conditional hooking followed by unconditional hooking is not applicable to the MSF algorithm. For example, the minimum outgoing edge $(i,j)$ from a vertex $i$'s star may not have $p^h_i < p_i$ so in that scenario, another edge would be used during conditional hooking. Instead, we require a function that operates on $p_i, a_{ij}, p_j$ all-at-once, which we discuss on Section \ref{ssec:mult_func}.

\section{Algebraic MSF}
\label{sec:algebraic_msf}
\subsection{Multilinear Function to Find Outgoing Edges} \label{ssec:mult_func}
We introduce a type of multilinear function that updates vertices by simultaneously using information from an edge and its two adjacent vertices. Consider the problem discussed in Section \ref{ssec:as} of computing the weight of the minimum outgoing edge from a vertex $i$ that belongs to a star. Let us denote the parent vector with $\bm{p}\in V^n$ where $p_i$ is the parent of vertex $i$. Since we assume that vertex $i$ belongs to a star, we decide whether an edge $(i,j)$ is outgoing from $i$'s star with $p_i \neq p_j$. We first seek to design a function $f$ that outputs $a_{ij}$ when edge $(i,j)$ is outgoing from $i$'s star and $\infty$ otherwise. Such functions are of the form $f:V\times \mathbb{R}\times V \to \mathbb{R}$, where
    \[  f(p_i,a_{ij},p_j) = \begin{cases}
          a_{ij} & : p_i \neq p_j \text{ and } i \text{ belongs to a star},\\ 
          \infty & : \text{otherwise}.
        \end{cases}\]
Notice that if the vertex $i$ did not belong to a star, then $f(p_i, a_{ij}, p_j) = \infty$ for any $a_{ij}$ and $p_j$. We may compute the weight of the minimum outgoing edge from $i$'s star with
\[
  q_i \underset{\min_j}{\gets} f(p_i, a_{ij}, p_j).
\]

More generally, this multilinear function is of the form $f:\mathbb{S}_{\bm{x}}\times \mathbb{S}_{\mat{A}}\times \mathbb{S}_{\bm{y}}\to \mathbb{S}_{\bm{w}}$, where
    \[w_i \gets \underset{j}{\bigoplus} f(x_i, a_{ij}, y_j).\]
with $\bm{w}\in \mathbb{S}^n_{\bm{w}}$, $\mat{A}\in \mathbb{S}^{n\times n}_{\mat{A}}, \bm{x}\in \mathbb{S}^n_{\bm{x}}, \bm{y}\in \mathbb{S}_{\bm{y}}^{n}$ and $\mathbb{S}_{\bm{w}}$ is equipped with a binary operation $\oplus$. Note that $f$ in the basic form $x_ia_{ij}y_j$ over the usual semiring $(\mathbb{R},+,*)$ is a bilinear function. In Section \ref{ssec:mult_kernel}, we analyze the complexity of evaluating multilinear functions of this type in terms of PRAM and communication cost.

\subsection{Algebraic MSF Algorithm}
In Algorithm \ref{alg:multilinear_mst}, we reformulate the AS algorithm using linear algebraic primitives. Given an undirected graph with distinct edge weights, the algorithm computes the weight of the MSF. For clarity of presentation, we omit tracking the MSF itself but note that we may do so storing by $(a_{ij}, i, j)$ in $\mat{A}$ and carrying this information throughout operations. We represent the parent forest with a parent vector $\bm{p}\in V^n$, where $p_i$ stores the parent of $i$. On line $1$, we define a set $\mathbb{EDGE}$ which contains pairs consisting of an edge weight and an entry from the parent vector. The monoid $(\mathbb{EDGE},\textsc{MinWeight})$ outputs the pair with the least edge weight.

We leverage the multilinear function described in Section \ref{ssec:mult_func} to compute the minimum outgoing edge from each vertex that belongs to a star on line $9$. Since we need the parent of the destination of the minimum outgoing edge to hook, we modify $f$ to return the pair $(a_{ij}, p_j)\in \mathbb{EDGE}$.
On line $10$, we project the minimum outgoing edges of the children onto their star root and keep the smallest such edge. We can write this projection more verbosely as
    \[
        r_i \gets \textsc{MinWeight}_j\{q_j : p_j=i\},
    \]
where $r\in \mathbb{EDGE}^n$. Intuitively, $r_i$ stores the minimum outgoing edge from the star with root $i$. Next, stars hook on line $11$. Note that only non-zero values of $\bm{r}$ are read, so $\bm{p}$ is unchanged for vertices that are not star roots. Since star hooking may create cycles in the parent forest, we detect which hooks must be removed to avoid cycles on line $12$. We use $\bm{t}$ to fix the parent forest and update $sum$ appropriately on lines $13$ and $14$. We finally shortcut on line $15$.



\setlength{\intextsep}{1\baselineskip}
\begin{algorithm}[!h]
\caption{Awerbuch-Shiloach with Linear Algebra}
\label{alg:multilinear_mst}
\begin{algorithmic}[1]
        \REQUIRE{$G=(V,E,w:E\to \mathbb{R})$ where $V=\{1,\dots,n\}$, is an undirected graph with distinct edge weights.}
        \STATE{We use the monoid $(\mathbb{EDGE},\textsc{MinWeight})$, where $\mathbb{EDGE} = \underbrace{\{w(e) : e \in E\} \cup \{\infty\}}_{\text{weight}} \times \underbrace{\{1,\ldots, n\}\cup \{0\}}_{\text{parent}}$ and given edges $x_1,\dots, x_k\in \mathbb{EDGE}$, $\textsc{MinWeight}\{x_1, \dots, x_k\}$ returns the edge $x_i$ with the least weight or $(\infty, 0)$ if $k=0$.}
        \newline
        \STATE{Let $\mat{A}\in \mathbb{R}^{n\times n}$ be the adjacency matrix of $G$, where\\
                $a_{ij} = \begin{cases} 
          w(i,j) & : (i,j) \in E  \\
          \infty & :  \text{otherwise}
        \end{cases}$}
        \STATE{Let $\vcr{p}^\text{old} = \begin{bmatrix} 0 & \cdots & 0 \end{bmatrix}^T$ be the old parent vector.}
        \STATE{Let $\vcr{p} = \begin{bmatrix} 1 & \cdots & n \end{bmatrix}^T$ be the parent vector.}
        \STATE{Let $f: V\times \mathbb{R}\times V \to \mathbb{EDGE}$, where\\
                $f(p_i,a_{ij},p_j) = \begin{cases}
                   (a_{ij}, p_j) & : p_i \neq p_j \\ 
                   (\infty, 0) & : \text{otherwise}
                 \end{cases}$}
        \STATE{$sum \gets 0$}
        \WHILE{$\vcr{p} \neq \vcr{p}^\text{old}$}
          \STATE{$\vcr{p}^\text{old} = \vcr{p}$}
          \STATE{$q_i \underset{\textsc{MinWeight}_j}{\gets} 
      \begin{cases}
        f(p_i, a_{ij}, p_j) & : i \text{ belongs to a star}\\
        (\infty, 0) & : \text{otherwise}
      \end{cases}$}
          \STATE{$r_{p_i} \underset{\textsc{MinWeight}_i}{\gets} q_i$}
          \STATE{$p_i\gets r_i^{parent}$}
          \STATE{$t_i \gets \begin{cases} \text{True} & : i \text{ is a star root and } i<p_i \text{ and } i=p_{p_i}\\
              \text{False} & : \text{otherwise} \end{cases}$}
              \STATE{$p_i \gets \begin{cases} i & : t_i = \text{True}\\
                  p_i & : \text{otherwise} \end{cases}$}
          \STATE{$sum \gets sum + 
                  \begin{cases} 
                    r_i^{\text{weight}} & : t_i = \text{False}\\
                    0 & : \text{otherwise}
                  \end{cases}$}
          \STATE{$p_i = 
      \begin{cases}
        p_{p_i} & : i \text{ does not belong to a star}\\
        0 & :  \text{otherwise}
      \end{cases}$}
        \ENDWHILE
        \STATE{$\text{return } sum$}
\end{algorithmic}
\end{algorithm}

\section{Parallel Analysis}
\label{sec:parallel_analysis}
\subsection{Multilinear Kernel} \label{ssec:mult_kernel}
We propose an all-at-once kernel to compute multilinear functions of the form introduced in Section \ref{ssec:mult_func}. We demonstrate that this multilinear kernel decreases the number of writes to main memory when compared to pairwise formulations.
We implement a multilinear kernel as a part of the CTF library and optimize the vector distribution compared to the default implementation.

\begin{figure}
\centering
\subfloat{\includegraphics[width = 2.8in]{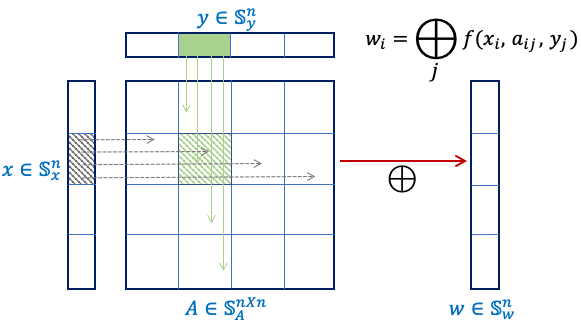}} 
\caption{Multilinear kernel communication pattern on a $2$D processor grid, where each processor owns a block of $\mat{A}$. We broadcast $\bm{x}$ over columns and $\bm{y}$ over rows. Each processor then computes $f(x_i,a_{ij},y_j)$ locally and contributes to the reduction $\bigoplus_j$ over columns.}
\label{fig:multilinear}
\vspace{-0.1in}
\end{figure}

\textbf{All-at-once:} We count the number of writes to main memory and the communication cost of this all-at-once kernel. We assume that $\mat{A}$ is mapped to a $2$D $\sqrt{p}\times \sqrt{p}$ processor grid. We denote processes on the grid by $(r,s)$ and use superscripts to denote the subset of an input assigned to each process. 
Also, we assume that the vectors $\bm{x},\bm{y},\bm{w}$ are partitioned along rows of the grid so that processor $(r,0)$ owns $\bm{x}^{(r)}$, $\bm{y}^{(r)}$, and $\bm{w}^{(r)}$.
First, we redistribute $\bm{y}$ to collect $\bm{y}^{(s)}$ along columns of the grid. Next, we broadcast $\bm{x}^{(r)}$ over all processes $(r,t)$ and $\bm{y}^{(s)}$ over all processes $(t,s)$, where $t$ is variable. Each process $(r,s)$ now owns $\mat{A}^{(r,s)}, \bm{x}^{(r)}, \bm{y}^{(s)}$ and computes $\bm{w}^{(r,s)} = f(\bm{x}^{(r)}, \mat{A}^{(r,s)}, \bm{y}^{(s)})$ locally. We then reduce over columns of the grid to yield $\bm{w}^{(r)} = \bigoplus_s \bm{w}^{(r,s)}$.
We see that this kernel requires $\frac{n}{\sqrt{p}}$ writes to main memory. 
The interprocess communication cost comprises of redistribution, broadcast, and reduction of a vector of local size $n/\sqrt{p}$. 
The amount of vector communication may in principle be reduced if the input vectors are sparse or if the output is sparse (or an output mask is supplied).
We visualize the data distribution and communication pattern in Figure \ref{fig:multilinear}.

\textbf{Pairwise:} Alternatively, this multilinear kernel can be implemented with two SpMV-like operations. However, we show that this approach requires $nnz$ more writes to main memory than the all-at-once approach. As an illustrative example, let us consider again the motivating problem for defining this type of multilinear function: given a vertex $i$ that belongs to a star, compute the weight of the minimum outgoing edge from $i$. Suppose we first compute a pairwise function $g(a_{ij}, p_j)$ which acts on $\mat{A}$ and $\bm{p}$. Since we have not yet determined which edges are outgoing from vertex $i$'s star with $p_i\neq p_j$, we update $\mat{A}$ to contain the pair $(a_{ij}, p_j)$. Now, we can redefine the function $f$ from Section \ref{ssec:mult_func} as a pairwise function $f(p_i, (a_{ij}, p_j))$ which acts on $\bm{p}$ and $\mat{A}$. This approach has a comparable communication cost as the all-at-once approach, but requires $nnz$ more writes to main memory. 

\subsection{Shortcutting}
\label{sec:optimizations}
For real-world graphs, we observe that it often takes only a few shortcut iterations to turn all trees into stars. We introduce complete shortcutting: at the end of each iteration, we shortcut repeatedly until each tree in the forest is a star. Since complete shortcutting allows each tree to attempt to hook at every iteration, at least half of the trees will hook successfully after the tie breaking step. However, complete shortcutting can perform at most $\log_{3/2} n$ shortcuts each iteration. While complete shortcutting requires fewer iterations than the original algorithm, the overall depth is worse by a $\log n$ factor. However, we observe in practice that the overhead of complete shortcutting is low and the tradeoff to converge in fewer iterations is worthwhile. In addition, complete shortcutting simplifies the AS algorithm by removing the need to query whether a vertex belongs to a star. To reduce the communication cost of complete shortcutting, we introduce an optimization called Complete Shortcutting with Prefetching (CSP).

The AS algorithm proposes the following shortcut step: for each vertex $i$ that does not belong to a star, update the parent vector $\bm{p}$ with 
 \[p_i \gets p_{p_i}.\]
Assuming that $\bm{p}$ is distributed on a vertex-based processor grid, we consider a baseline implementation that queries $p_{p_i}$ in main memory for each vertex $i$ assigned to this process that does not belong to a star and then performs the update $p_i \gets p_{p_i}$ locally. We collect duplicate queries from a single process to avoid querying $p_{p_i}$ multiple times for vertices with the same parent.

If the number of vertices whose parent changed after hooking is small, we may collect them on all processes and use only local data for complete shortcutting. Let us denote the parent vector before the hooking step with $\bm{p}^{prev}$. We compare our local data of $\bm{p}$ and $\bm{p}^{prev}$ to determine which vertices have a new parent after the hooking and tie-breaking steps on lines $2-6$. Notice that a vertex $i$ has a new parent if and only if $i$ is a star root that successfully hooked onto another star. On line $7$, we collect all such vertices and their new parents on all processes. We then implement complete shortcutting using only local data on lines $8-12$. If the parent of a vertex is a star root that did not hook onto another star, we stop shortcutting. We provide a detailed description of CSP in Algorithm \ref{alg:shortcut3}.


\begin{algorithm}
\caption{Complete Shortcutting with Prefetching}
\label{alg:shortcut3}
\begin{algorithmic}[1]
    \REQUIRE{$\bm{p}\in V^n \text{ is the parent vector},$ \\  $\bm{p}^{prev}\in V^n \text{ is the previous parent vector}.$} 
    \STATE{$changed \gets \varnothing$}
    \COMMENT{$changed$ is a map}
    \FOR{$(i, p_i) \text{ in local data}$}
      \IF{$p_i \neq p^{prev}_i$}
        \STATE{$changed\gets \{(i, p_i)\} \cup changed$}
      \ENDIF
    \ENDFOR
    \STATE{$changed \gets \operatorname{Allgather}(changed)$}
    \FOR{$(i, p_i) \text{ in local data}$}
      \WHILE[local data]{$p_i \text{ in } changed$}
      \STATE{$p_i \gets changed[p_i]$}
      \ENDWHILE
    \ENDFOR
    \STATE{$\text{return } \bm{p}$}
\end{algorithmic}
\end{algorithm}




\section{Related Work}
\label{sec:related_work}

%

\subsection{PRAM Algorithms}
The randomized linear time MST algorithm by Klein et al. \cite{linear-work} inspired a search for linear-work PRAM algorithms. First, Cole et al. proposed such an algorithm \cite{Cole94alinear-work} in the concurrent read, concurrent write (CRCW) model. Pettie and Ramachandran then developed a logarithmic depth and linear work algorithm \cite{Pettie99arandomized} in the exclusive read, exclusive write (EREW) PRAM model.

\vspace{-2mm}
\subsection{Parallel Implementations}
\label{ssec:parimpl}
Many shared and distributed memory graph frameworks provide an implementation for MST. Galois \cite{galois-shared} provides a shared memory implementation of MST using constructs defined in its programming model. To process large graphs on a single machine (with enough memory), Dhulipala et al. develop various scalable graph algorithms \cite{gbbs} including MST. Their approach in many cases is shown to outperform the distributed memory implementations. GraphChi \cite{graphchi} is another single machine implementation that can process massive graphs from secondary storage.

STAPL \cite{stapl} provides support for both shared and distributed memory parallelism in C++. MST can be implemented in STAPL using the framework's distributed data structures and parallel algorithms. Pregel \cite{pregel} uses a Bulk Synchronous Parallel (BSP) model, and takes a vertex centric approach for graph computations. 
The API provided by the framework can be used to program many graph algorithms in a distributed environment. Data distribution, underlying message handlers, and fault tolerance are invisible to the user. GPS \cite{gps} extends the Pregel API to incorporate dynamic repartitioning, among other optimizations. 

One of the shortcomings of Pregel's vertex centric approach is the message load imbalance caused by few vertices that communicate more messages than others. To address the shortcomings, Pregel+ \cite{pregel_plus} proposes two techniques, vertex-mirroring and request-respond paradigm where high degree vertices are mirrored, and all requests from a machine to the same target are merged into one request. Pregel+ is shown to outperform other distributed memory frameworks including GPS \cite{gps} and Powergraph \cite{powergraph}. An experimental evaluation that compares various graph frameworks (across different algorithm categories, graph characteristics, etc.) notes that there is no single system that has superior performance in all cases \cite{pregel_comparison}.

Panja et al. propose MND-MST to compute MST on heterogeneous systems that house both CPU and GPU compute capabilities \cite{mnd-mst}. They partition the input graph across multiple nodes and devices and compute local MSTs in parallel using Boruvka's algorithm. They employ a 1D partitioning scheme to balance the number of edges across computing units. They present results for both CPU-only and multi-device (CPU-GPU) systems. Their results are shown to outperform Pregel+. We qualitatively compare Pregel+ and MND-MST to our MSF implementation in Section \ref{sec:pregel_mnd}.






GraphBLAS \cite{graphblas} provides standardized linear-algebraic primitives for graph computations. LAGraph \cite{lagraph} builds on top of GraphBLAS to provide developers of graph algorithms a set of data structures and utility functions. They describe and benchmark algebraic implementation of several graph algorithms using LAGraph's API, though not for MST.
\subsection{LACC and FastSV}
We further compare and contrast our MSF formulation with the previously proposed linear algebraic frameworks for graph connectivity, LACC \cite{lacc} and FastSV \cite{fastsv}. 
Both perform similar hooking and shortcutting steps. The formulations identify sets of active vertices that contribute to the output of certain computations and represent them with a sparse vector. For example, vertices that belong to converged components are inactive.


If the parent vector $\bm{p}$ has converged, we still need a last iteration to verify that. FastSV proposes a stronger termination condition: repeat until convergence of the grandparent vector. Since both CC and MSF terminate when a spanning forest is found, this condition holds for MSF as well. For most real-world graphs, the last iteration does not perform any hooking and only shortcuts trees into stars. In these cases, the stronger termination condition identifies a spanning forest an iteration before all trees are shortcut into stars. 

FastSV proposes three hooking optimizations: hooking onto a grandparent, stochastic hooking, and aggressive hooking. Hooking onto a grandparent  results in shorter trees after the hooking step. However, this optimization is not applicable with complete shortcutting. The second optimization, stochastic hooking, relaxes the hooking condition to allow hookings to happen more often. Intuitively, stochastic hooking allows a tree to be split into multiple parts, each of which hook independently. However, this would violate the requirement for MST that we hook using the $\textit{minimum}$ outgoing edge from a star. 
The last hooking optimization, aggressive hooking, would also violate this requirement.
\vspace{-2mm}
\subsection{All-at-once Kernels}
For routines like the tensor-times-tensor-product (TTTP) that arise in tensor completion, Singh et al. \cite{tttp} propose a multi-tensor all-at-once contraction that replaces the standard approach of pairwise contraction. We follow a similar approach in the development of our multilinear kernel. In a very recent work~\cite{fusedmm}, Rahman et al. introduce a kernel called FusedMM that generalizes sampled dense-dense matrix multiplication and sparse-dense matrix multiplication. Our multilinear kernel can be viewed as an instance of FusedMM, where the message generated on each edge is the weight of the edge. While our multilinear kernel is motivated by MSF, FusedMM is motivated by graph embeddings and graph neural networks.

\section{Experimental Setup}
We evaluate the performance of our implementation on the Stampede2 supercomputer. Each node has an Intel Xeon Phi 7250 CPU (``Knights Landing") with 68 cores, 96GB of DDR4 RAM, and 16GB of high-speed on-chip MCDRAM memory (which operates as 16GB direct-mapped L3).
In our experiments, we use 16 MPI processes each with 4 OMP threads per node.
We use Cyclops Tensor Framework (CTF, on raghavendrak fork v1.5.6) for our implementation. CTF supports a variety of generalized vector/matrix/tensor operations. We mainly rely on the data distribution and algebraic primitives supported by CTF. We optimize vector distribution/transpose operations, and use them to implement the multilinear kernel. The implementation is generic, and can be used in the development of other algorithms that use multilinear primitives. Our code is available at https://github.com/raghavendrak/algebraic\_MSF.

\begin{table}[htbp]
  \begin{adjustbox}{width=\columnwidth,center}
    \begin{tabular}{c|c|c|c}
      \textbf{ID} & \textbf{Graph} & \textbf{Nodes} & \textbf{Edges} \\
      \hline
      friendster & Friendster & 65.6M & 1.8B \\
      \hline
      orkut & Orkut social network & 3.1M & 117.2M \\
      \hline
      lj & LiveJournal social network & 4M & 34.7M \\
      \hline
      road-usa & Full USA road-network & 23.9M & 28.9M \\
      \hline
      road-central & Central USA road-network & 14.1M & 16.9M \\
      \hline
      agatha\_2015 & Deep-learning graph & 183.9M & 11.6B \\
      \hline
      moliere\_2016 & Hypothesis generation network & 30.2M & 6.7B \\
      \hline
    \end{tabular}
  \end{adjustbox}
    \caption{\label{tab:graphs} \textbf{Real-world undirected graphs}}
\vspace{-0.2in}
\end{table}

\section{Evaluation}
\label{sec:evaluation}
We present performance results for our algebraic implementation of MSF using the following classes of input graphs:
\begin{itemize}
  \setlength\itemsep{0.05em}
  \item road-network graphs from the DIMACS graph partitioning and clustering challenge \cite{dimacs},
  \item real-world social-network graphs from the SNAP dataset \cite{snapnets},
  \item two of the largest graphs in terms of number of edges from SuiteSparse matrix collection \cite{tamu},
  \item synthetic uniform random and R-MAT graphs \cite{r-mat}.
\end{itemize}
The various graphs used in our evaluation are listed in Table \ref{tab:graphs}. 
For example, the MST of the full USA road-network is the minimum weight set of roads that connect cities in the USA.
We use both R-MAT and real-world graphs to show strong scaling results. We use uniform random graphs to present weak scaling results. 
For unweighted graphs, we generate uniformly distributed integers in $[1,255]$ for edge weights. Note that while the MSF for such graphs may not be unique, our algorithm outputs an optimal MSF.
This choice is consistent with previous performance studies of graph algorithms such as the GAP benchmark suite \cite{gap} and Graph 500's SSSP proposal \cite{graph500}. We report \textit{runtime} as the execution time of our MSF implementation excluding graph I/O.
\subsection{Shortcut optimization}  
\begin{figure}[t]
\centering
\includegraphics[width = 2.4in]{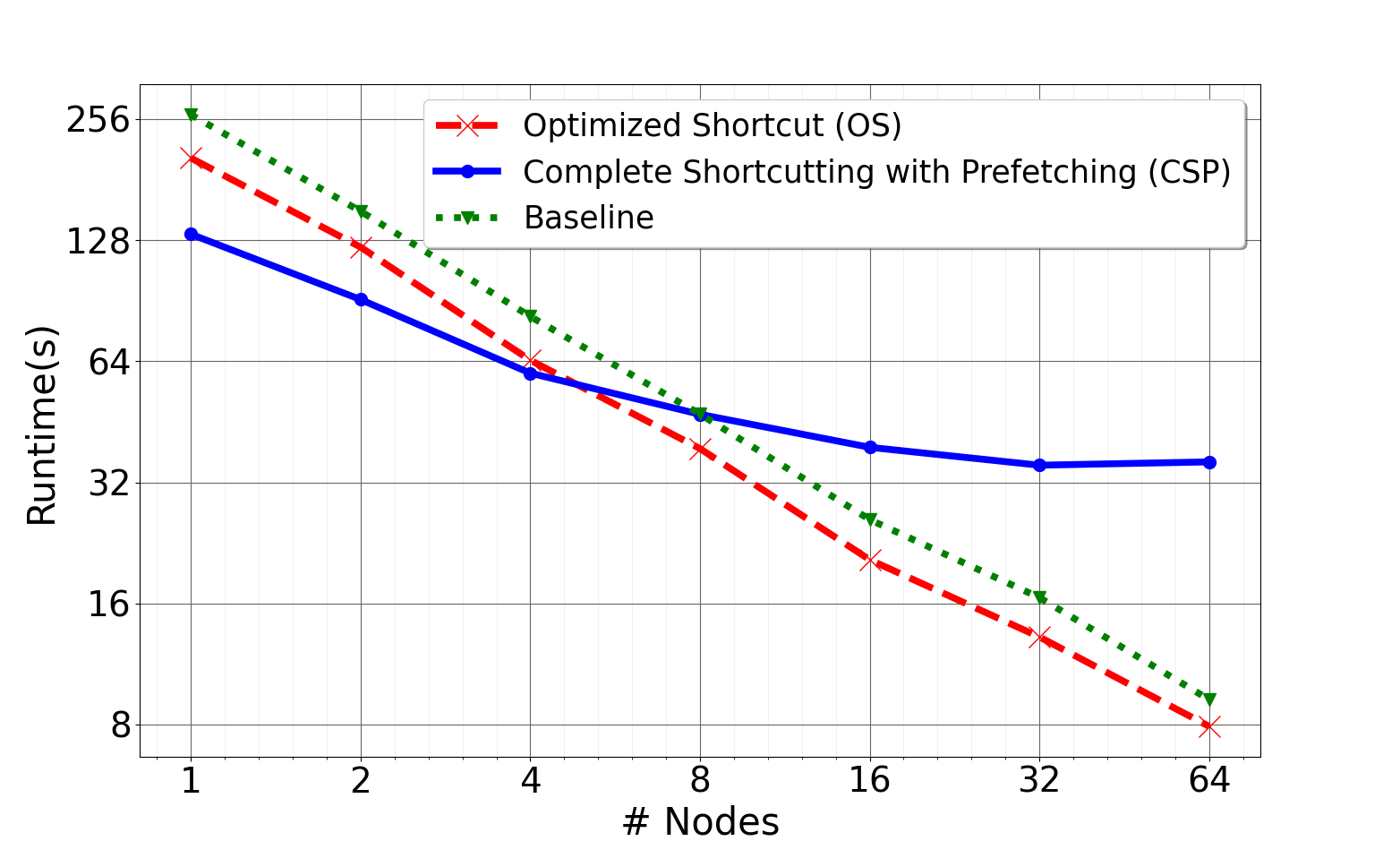}
\caption{Performance of our MSF implementation with shortcut optimizations on the full USA road-network.}
\label{fig:compare_st}
\vspace{-0.2in}
\end{figure}

\begin{figure}[t]
\centering
\includegraphics[width = 2.4in]{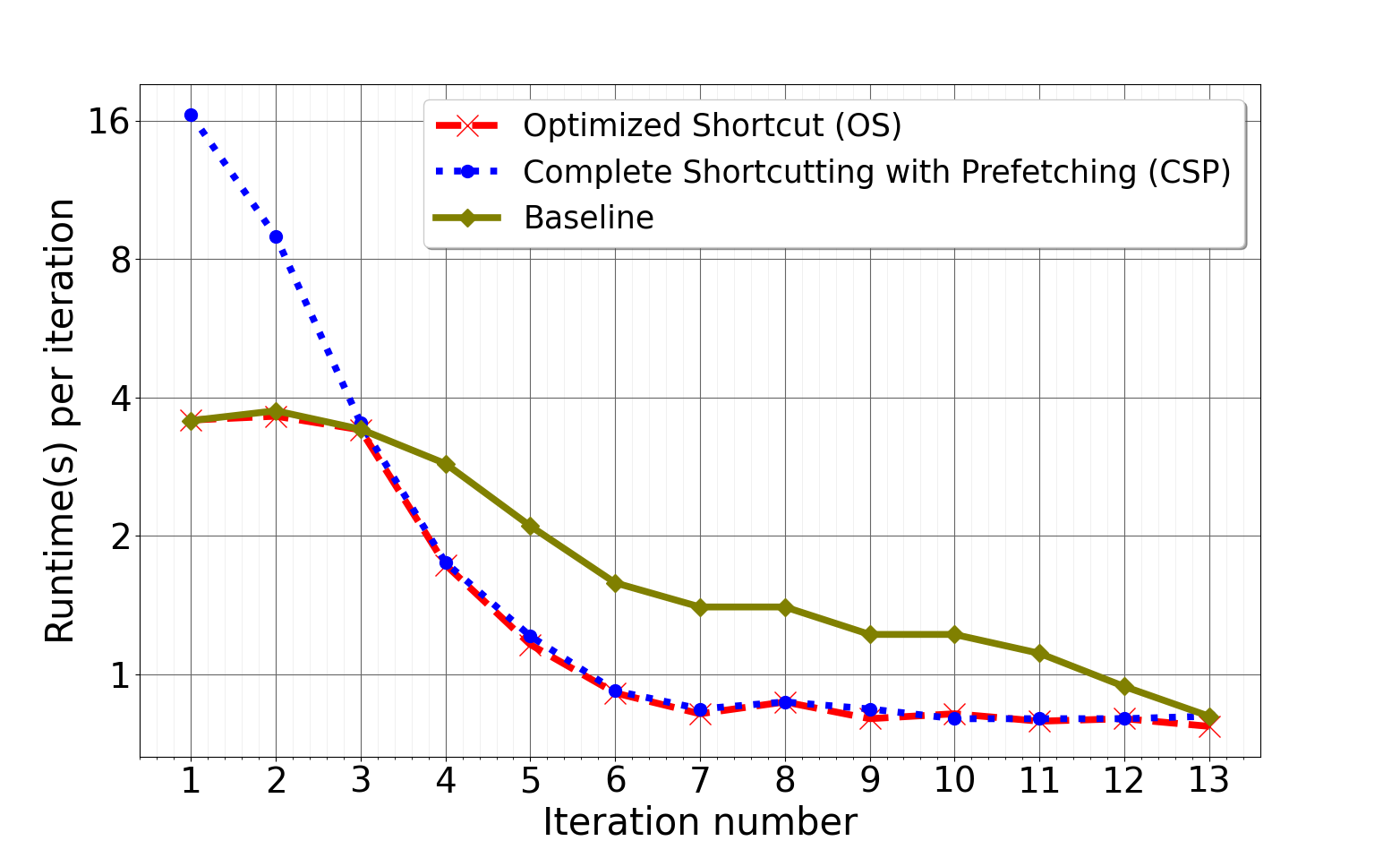}
\caption{Performance per iteration of our MSF implementation with shortcut optimizations on the full USA road-network with 16 nodes.\\}
\label{fig:st3}
\vspace{-8mm}
\end{figure}




In Figure \ref{fig:compare_st}, we compare the performance of our MSF implementation with the shortcutting optimizations proposed in Section \ref{sec:optimizations}.
A single invocation of complete shortcutting performs several iterations to convert every tree in the forest to a star. We refer to these shortcutting iterations as sub-iterations.
Many vertices in a process might have the same parent, and hence a process might send duplicate queries to the parent vector. In the baseline implementation we do not optimize the shortcut function except for collecting duplicate queries from a single process. The CSP optimization as described in Section \ref{sec:optimizations} performs complete shortcutting using only local data after a collective communication step.
Optimized Shortcutting (OS) is a combination of CSP and the baseline implementation. In OS, we invoke CSP only if the number of vertices to be gathered is below a fixed threshold. Otherwise, the baseline implementation is invoked. We use an empirically determined threshold of 1310k, which translates to 20MB gathered on every process.
We present data for runs where the respective optimizations are invoked in every iteration.


For up to 4 nodes, invoking CSP in all iterations performs the best. The predominant step in the baseline implementation is the parent vector query performed every sub-iteration. CSP on the other hand gathers data only once during complete shortcutting. 
The parent vector is distributed across nodes. With the increase in node count, since each process holds fewer vector elements, the query needs to fetch fewer elements per process. We see that it is beneficial to have no optimization in shortcut for the first few sub-iterations.
If CSP has to gather substantial amount of data, the benefit of skipping the reads in every sub-iteration is lost. 
CSP performs better at lower node counts or when the amount of data gathered is small. 
The algorithm computes MST for road\_usa in 13 iterations. 
As the algorithm progresses, fewer vertices hook on to new parents so the threshold is crossed from the fourth iteration.

Figure \ref{fig:st3} compares the runtime per iteration of our MSF implementation with different shortcutting optimizations on 16 nodes. CSP outperforms the baseline implementation from the fourth iteration but is relatively expensive in the first two iterations. The optimized shortcut invokes CSP only after three iterations of the algorithm when fewer vertices hook on to a new parent. For certain graphs we observe that the switch to CSP's shortcut is not always triggered at an optimal point, i.e. an earlier switch to CSP could have achieved better overall performance. A dynamic threshold would further improve performance.


\begin{figure*}[t]
\subfloat[Social network graphs]{\includegraphics[width = 2.4in]{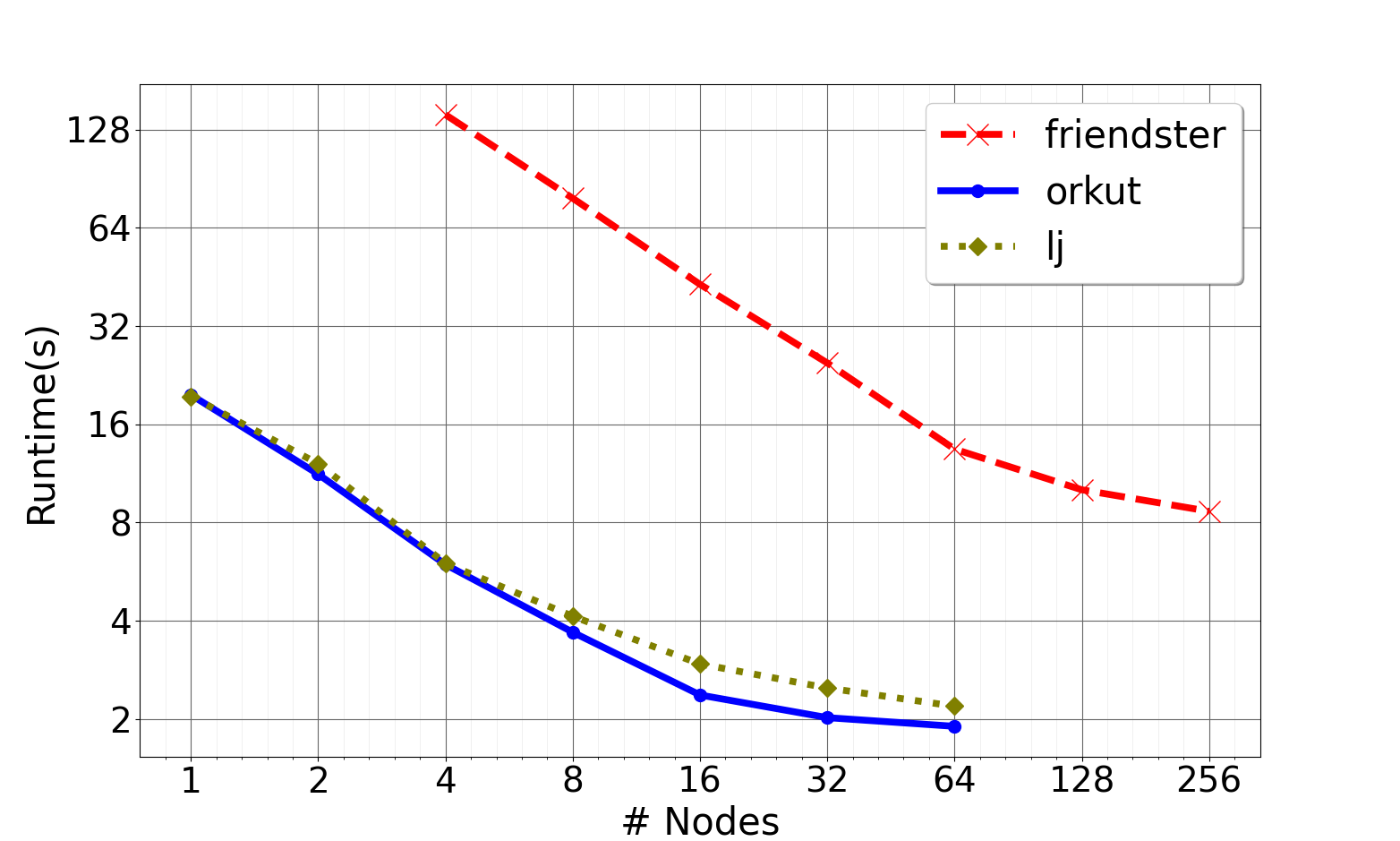}} 
\subfloat[DIMACS road network graphs]{\includegraphics[width = 2.4in]{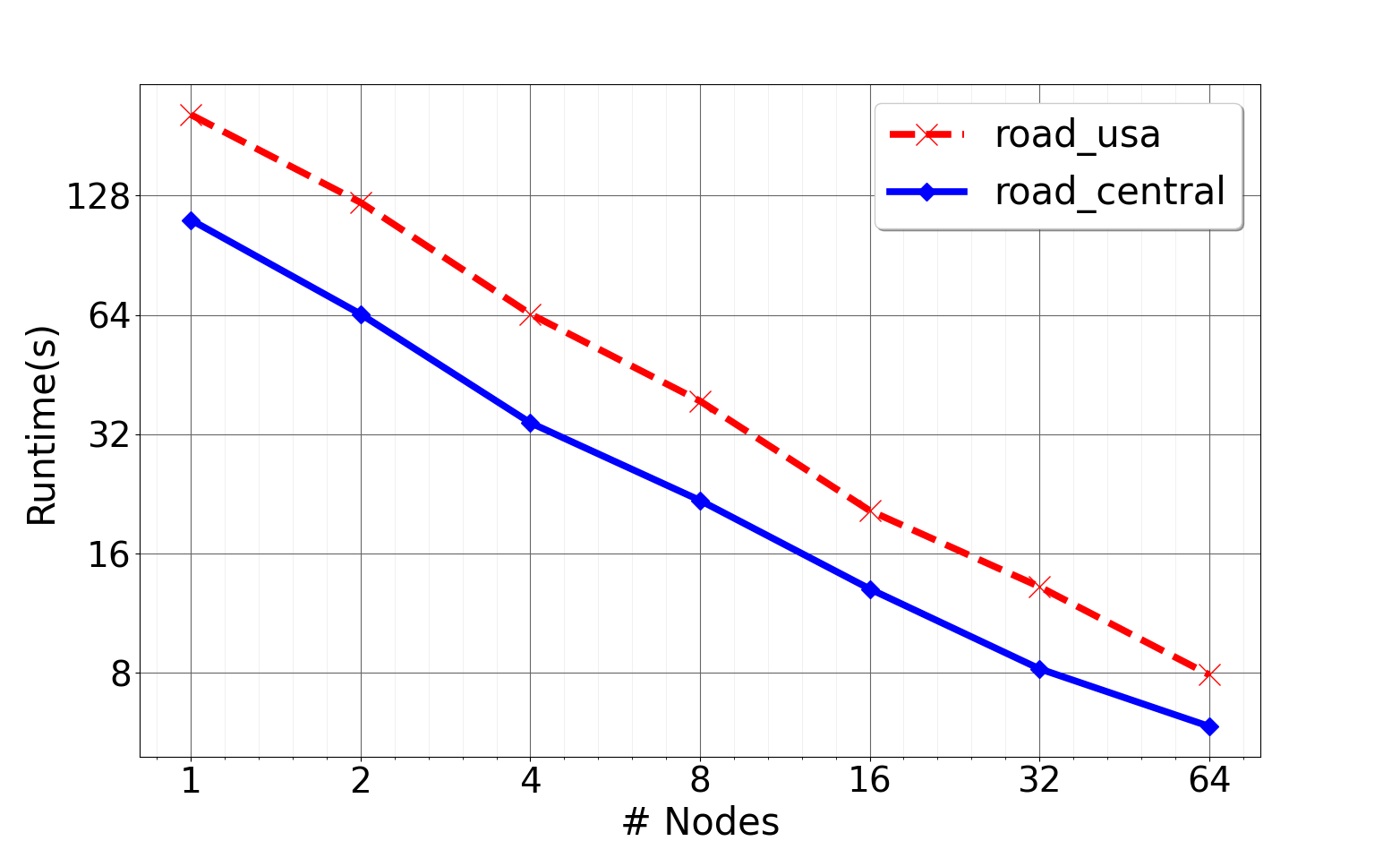}}
\subfloat[SuiteSparse graphs]{\includegraphics[width = 2.4in]{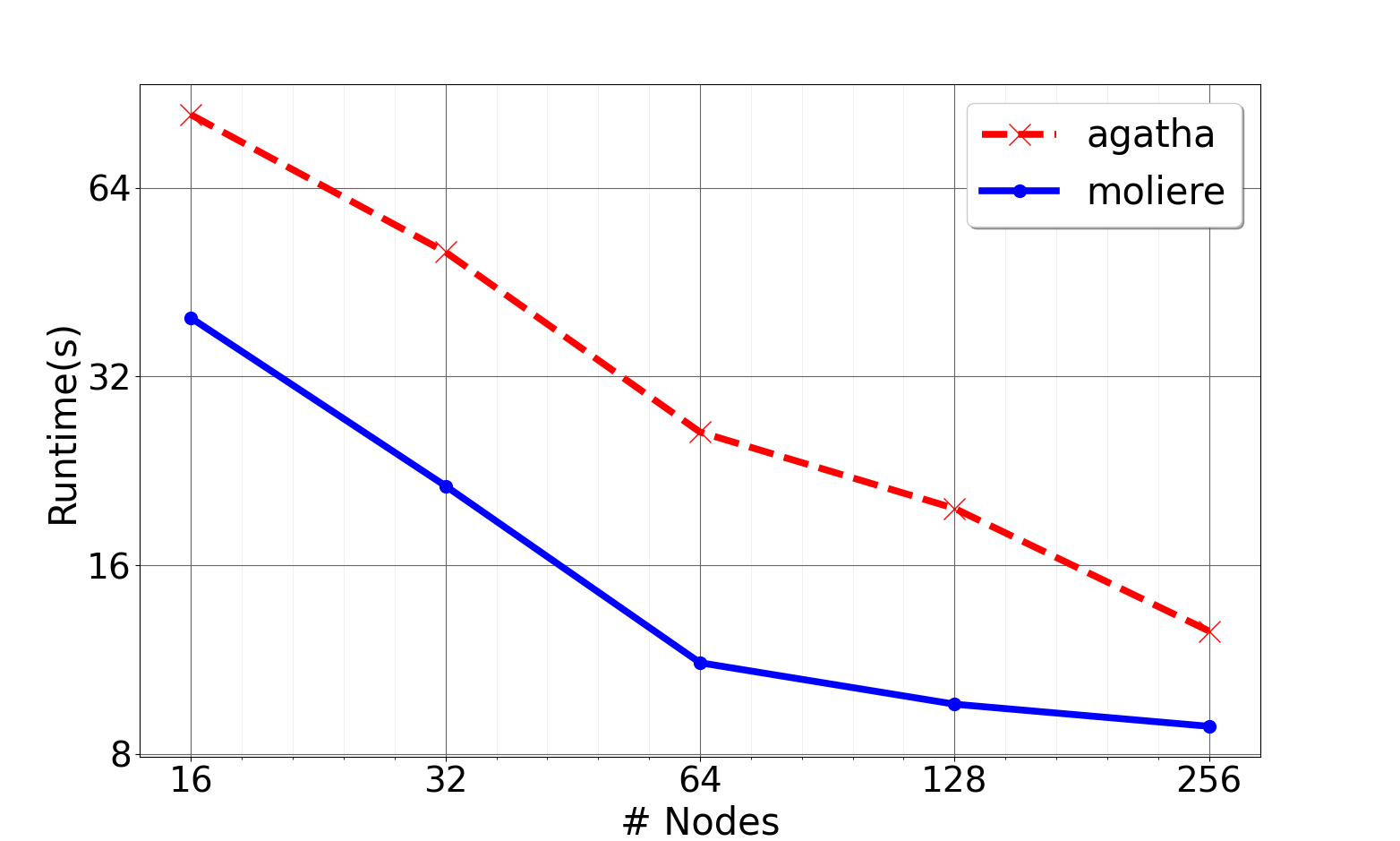}}\\
\caption{Strong scaling of real-world graphs.}
\label{fig:real_world_graphs}
\vspace{-0.3in}
\end{figure*}

\subsection{Strong Scaling}
We study the performance of our implementation with CSP as the node count increases. We consider three social-network graphs in Figure \ref{fig:real_world_graphs}(a). Orkut and LiveJournal show similar scaling results. Orkut is relatively denser but has a smaller diameter when compared to Friendster and LiveJournal. The time per iteration is higher in Orkut, but the number of iterations for the algorithm to converge is higher by one for LiveJournal. On a single node, the multilinear kernel for Orkut takes 1.41s whereas for LiveJournal it is 0.65s, per iteration. For a fixed node count, the runtime for each call of the multilinear kernel is fairly constant across iterations.

Friendster is the largest graph available in the SNAP dataset. 
The smallest number of nodes that can compute MSF for Friendster successfully is 4. Given the extremely large number of vertices and edges, both the computation of the multilinear kernel and the communication in the shortcut step are expensive. 
We note that the runtime for a call of the multilinear kernel achieves good scalability for all the three graphs. However, the overall time per iteration reduces as the algorithm progresses since fewer vector elements need to be updated during shortcutting.

From the DIMACS dataset~\cite{dimacs} we use the road-network graphs in Figure \ref{fig:real_world_graphs}(b). These graphs have a larger diameter and are sparser than the social network graphs. On a single node, while the computation time for the multilinear kernel is comparable to that of the Orkut graph, the number of iterations required to compute the MST is roughly doubled. These graphs have 5X to 8X more vertices than Orkut, so the runtime per iteration is higher due to complete shortcutting being more expensive. We achieve good strong scaling until 64 nodes with around 26X speedup over 1 node.

In Figure \ref{fig:real_world_graphs}(c), we present strong scaling results for the two largest graphs from the SuiteSparse matrix collection, both of which are dense. The machine learning graph Agatha has around 700X more edges than \textit{road-central}, the smallest graph considered until now. The minimum number of nodes required to finish is 16. We achieve good scaling until 256 nodes, which is the maximum number of nodes that we tried.

\begin{figure}[t]
\centering
\includegraphics[width = 2.4in]{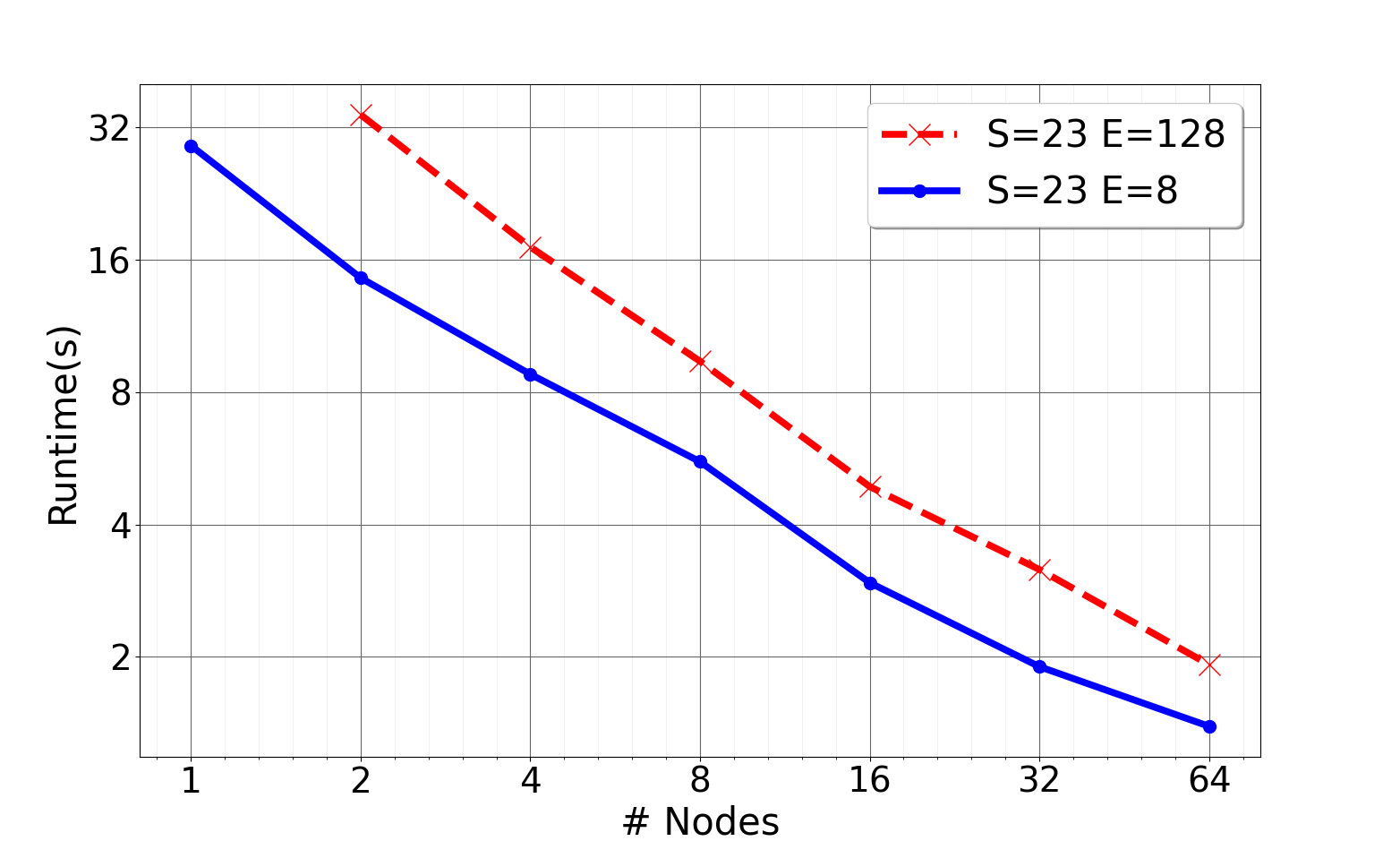}
\caption{Strong scaling of R-MAT graph}
\label{fig:rmat_graphs}
\vspace{-0.2in}
\end{figure}

\textbf{R-MAT graphs:} We use two R-MAT graphs with $\log_2(n) \approx S = 23$, and average degree controlled by $k \approx E ={8, 128}$. In Figure \ref{fig:rmat_graphs} we show the strong scaling results for the two graphs. Both the graphs require only four iterations to compute MST. The computation time for shortcutting is fairly comparable for both the graphs. Since the graph with $E=8$ is sparse when compared to the graph with $E=128$, the multilinear kernel is only a fraction of the iteration time. When $E=128$ the multilinear kernel computation dominates the overall iteration time. Both graphs show good strong scalability until 64 nodes. 
%

\subsection{Weak Scaling}
\begin{figure}[t]
\centering
\includegraphics[width = 2.4in]{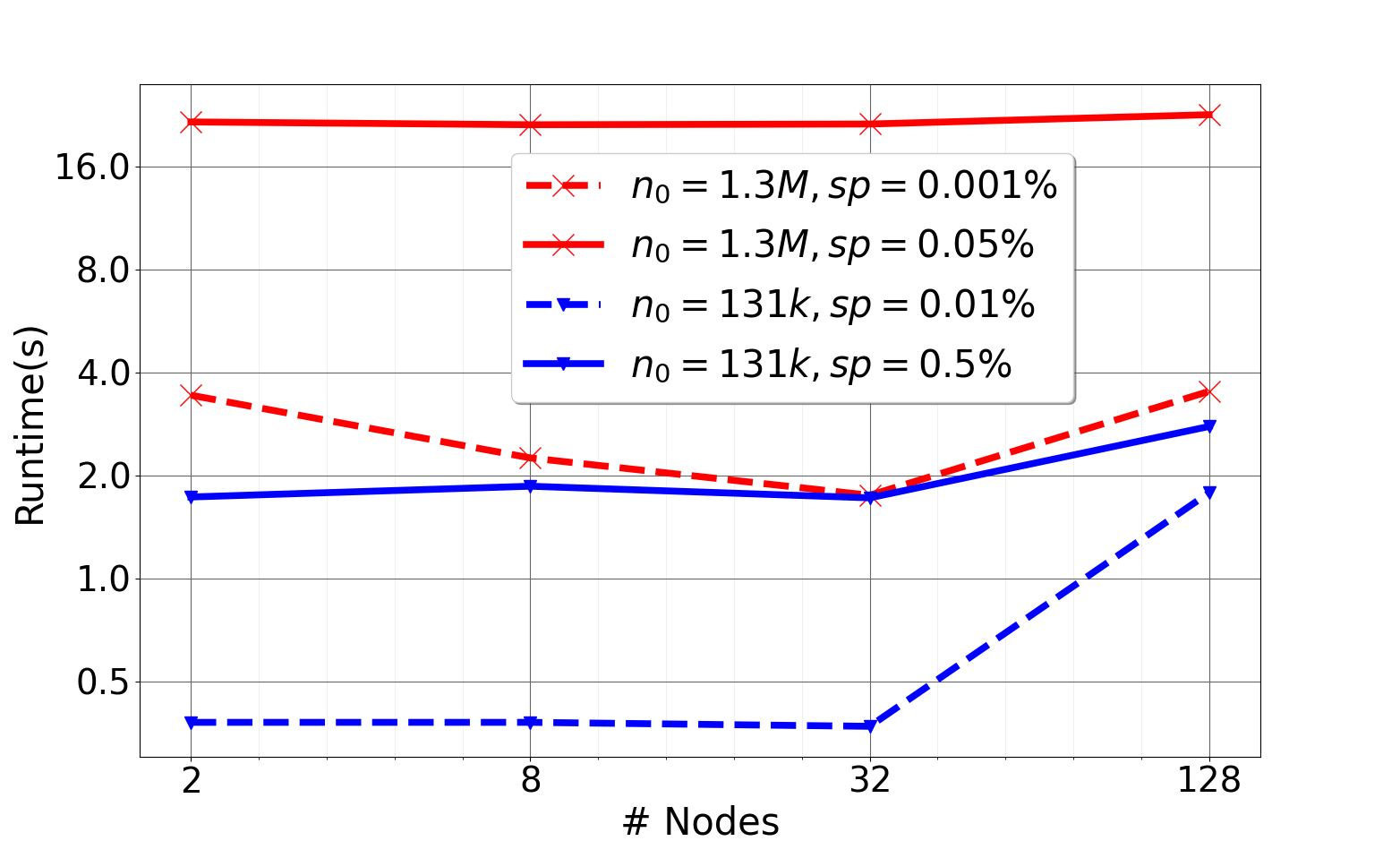}
\caption{Edge weak scaling of uniform random graphs. Constant $n^2 / p = n_{0}^2$ and edge percentage $f = 100 * m / n^2$.}
\label{fig:uniform_random_graphs}
\vspace{-0.2in}
\end{figure}  
We use uniform random graphs, in which all nodes have the same expected degree to study weak scaling. We consider ``edge weak scaling'' where $n^2/p$ is kept constant. We achieve good weak scaling as shown in Figure \ref{fig:uniform_random_graphs}. The number of iterations is smaller when the sparsity is $0.05\%$ or $0.5\%$ when compared to sparsity of $0.001\%$ or $0.01\%$, but the multilinear computation time per iteration is the most significant factor for dense graphs.

\begin{figure}[t]
\centering
\includegraphics[width = 2.4in]{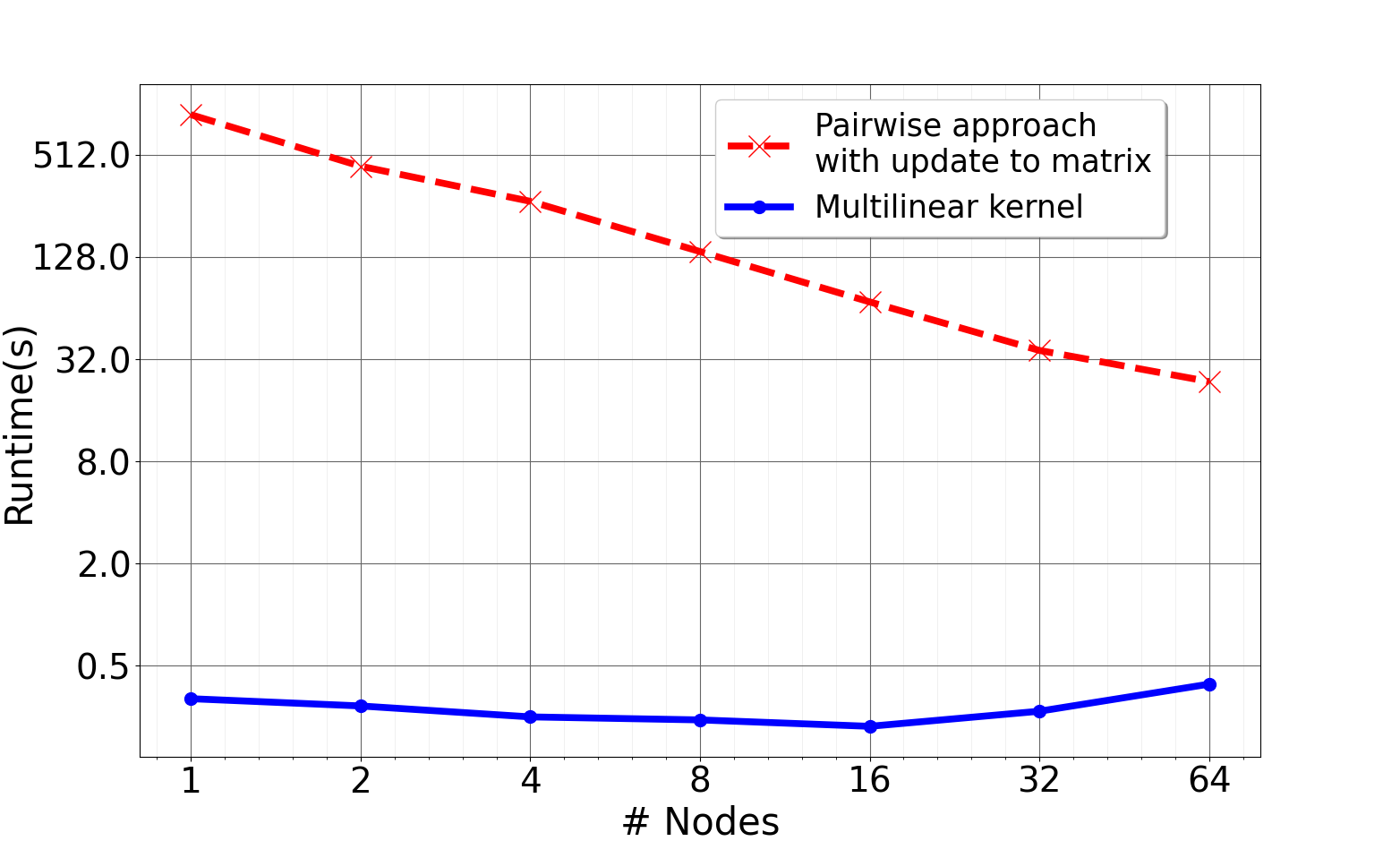}
\caption{A case for multilinear kernel on R-MAT graph with $S=17$ and $E=8$.}
\label{fig:as_multilinear}
\vspace{-0.3in}
\end{figure}

In Figure \ref{fig:as_multilinear}, we evaluate performance on an R-MAT graph with and without using the multilinear kernel. The multilinear kernel implementation outperforms the pairwise implementation described in Section \ref{ssec:mult_kernel} by several orders of magnitude. 
\vspace{-2mm}
\subsection{Comparison with Pregel+ and MND-MST}
\label{sec:pregel_mnd}
We discuss the approach and optimizations introduced by Pregel+~\cite{pregel_plus} and MND-MST~\cite{mnd-mst} in Section \ref{ssec:parimpl}.
Both performance studies use 16 nodes and present results for the \textit{road\_usa} graph. MND-MST discusses scaling results, while Pregel+ does not.
We note that the computing platforms are different in each of the experimental setups. Pregel+ reports 19.95s to compute MST for \textit{road\_usa}. 
On the same graph, Panja et al. reports 190s and 29.6s using Pregel+ and MND-MST, respectively. They report slowdowns beyond 16 nodes and note that the communication overhead is higher at larger node counts. We show strong scalability for \textit{road\_usa} with an execution time of 20.6s and 7.9s on 16 and 64 nodes, respectively.
While the architectures and backend systems are different (Pregel+ uses Hadoop, which is not a standard module in Stampede2), these results show that our algebraic MST implementation is roughly comparable in performance to hand-optimized MST codes.



\section{Conclusions}

Multi-tensor contraction kernels have shown to be advantageous in tensor completion operations such as TTTP and MTTKRP. We show that a similar all-at-once approach in the algebraic formulation for MSF is efficient when compared to a pairwise approach which involves an expensive update.
We suspect that such multilinear kernels or their simple variants can be leveraged to optimize existing algebraic algorithms and design new ones.

We observe that the number of shortcuts required for complete shortcutting in practice is much lower than upper bound discussed.
Complete shortcutting simplifies the AS algorithm by removing the need to query whether a vertex belongs to a star.
We show that our algebraic formulation with the CSP optimization achieves excellent strong and weak scaling for various graphs including some of the largest available real-world graphs.

\bibliographystyle{siam}
\bibliography{paper}

\end{document}